\documentclass[aps,twocolumn,superscriptaddress,longbibliography]{revtex4-1}

\usepackage[colorlinks=true,citecolor=blue]{hyperref}
\usepackage{graphicx}
\usepackage{bm}
\usepackage{amsmath}
\usepackage{amssymb}

\DeclareGraphicsExtensions{.png,.jpg,.eps}
\usepackage{xcolor}

\newcommand{\be}{\begin{equation}}
\newcommand{\ee}{\end{equation}}
\newcommand{\beq}{\begin{eqnarray}}
\newcommand{\eeq}{\end{eqnarray}}
\begin{document}

\author{Guangze Chen}
\affiliation{Department of Applied Physics, Aalto University, 02150 Espoo, Finland}

\author{Maryam Khosravian}
\affiliation{Department of Applied Physics, Aalto University, 02150 Espoo, Finland}

\author{Jose L. Lado}
\affiliation{Department of Applied Physics, Aalto University, 02150 Espoo, Finland}

\author{Aline Ramires}
\affiliation{Condensed Matter Theory Group, Paul Scherrer Institute, CH-5232 Villigen PSI, Switzerland}

\title{Designing spin-textured flat bands in twisted graphene multilayers via helimagnet encapsulation}

\begin{abstract}

Twisted graphene multilayers provide tunable platforms to engineer flat bands and exploit the associated strongly correlated physics. The two-dimensional nature of these systems makes them suitable for encapsulation by materials that break specific symmetries. In this context, recently discovered two-dimensional
helimagnets, such as the multiferroic monolayer NiI$_2$, are specially appealing for breaking time-reversal and inversion symmetries due to their nontrivial spin textures. Here we show that this spin texture can be imprinted on the electronic structure of twisted bilayer graphene by proximity effect. We discuss the dependence of the imprinted spin texture on the wave-vector of the helical structure, and on the strength of the effective local exchange field. Based on these results we discuss the nature of the superconducting instabilities that can take place in helimagnet encapsulated twisted bilayer graphene. Our results put forward helimagnetic encapsulation as a powerful
way of designing spin-textured flat band systems, providing a starting point to engineer a new family of correlated moire states.
\end{abstract}

\date{\today}

\maketitle

\section{Introduction}

Magnetic van der Waals materials have become a fundamental building block in artificial 
heterostructures\cite{huang2017layer,Gibertini2019}. 
Their two-dimensional nature
provides a versatile platform to electrically control magnetism\cite{Deng2018,Huang2018,Jiang2018,PhysRevLett.117.267203}, 
design magnetic tunnel junctions\cite{ghazaryan2018magnon,Klein1218,Song2018,jiang2018spin},
create artificial magnetic structures\cite{PhysRevResearch.3.013027,Akram2021,2021arXiv210309850X},
and topological superconductivity\cite{Kezilebieke2020,2020arXiv201109760K}.  
Magnetic encapsulation with van der Waals materials further provides a knob
to engineer artificial quantum states\cite{PhysRevLett.121.067701,Sierra2021,Soriano2020,Mashhadi2019,PhysRevLett.123.237201,PhysRevLett.126.056803,2021arXiv210207484C,PhysRevB.100.085128,Soriano2021}.
Van der Waals helimagnets, as realized in transitional metal dihalides\cite{McGuire2017,PhysRevMaterials.3.044001,PhysRevB.30.2140,Amoroso2020},
provide a direction to exploit non-collinear magnetism in van der Waals
heterostructures.
In particular, non-collinear magnetism associated to two-dimensional multiferroics
such as NiI$_2$\cite{2021arXiv210607661S} can potentially lead to new strategies to control van der Waals magnetism electrically\cite{Hill2000,Ramesh2007,Fiebig2016}. 

Twisted graphene multilayers have become
paradigmatic systems to engineer flat bands,
leading to a variety of unconventional many-body states\cite{PhysRevB.82.121407,Bistritzer2011,Cao2018,Cao2018super,Lu2019,Yankowitz2019,PhysRevLett.124.076801,PhysRevB.99.235406,PhysRevLett.126.026801,PhysRevLett.124.106803,Serlin2019,PhysRevLett.127.026401,PhysRevB.102.115127,Can2021}.
The role of encapsulation in twisted graphene multilayers, in particular with boron nitride,
has been shown to have a critical role in promoting
specific correlated states such as Chern insulating states in twisted bilayers\cite{Serlin2019}
and superconductivity in graphene trilayers\cite{2021arXiv210607640Z}.
Spin-orbit coupling proximity effects have also been exploited to promote
specific correlated states\cite{2021arXiv210206566L,2021arXiv210812689B}, and its combination with magnetic encapsulation
has been shown to promote unconventional symmetry broken states\cite{PhysRevLett.126.056803}.
However, proximity effect to non-collinear van der Waals
materials has remained unexplored, especially as a means of controlling
twisted van der Waals quantum states.

\begin{figure}[t!]
\center
\includegraphics[width=\linewidth]{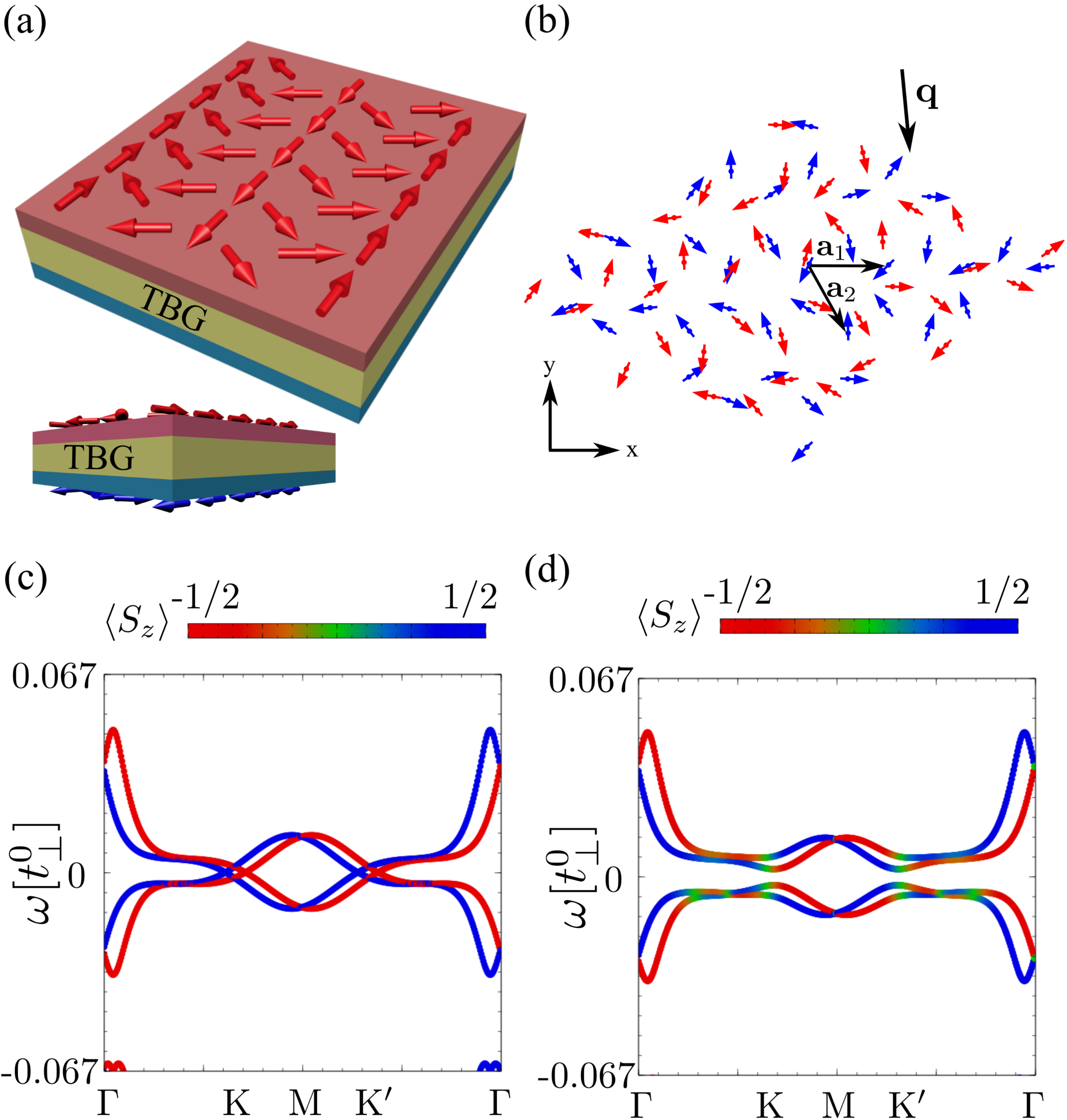}
\caption{(a) Sketch of TBG encapsulated with helimagnets, viewed from top and side. (b) Local effective exchange field induced by proximity to helimagnets on top (red) and bottom (blue) layers of TBG. The direction of characteristic vector of the helimagnets $\mathbf{q}$ is denoted with the black arrow. (c) Band structure of the heterostructure described by Eq. \eqref{eq6} with $J_0=0$ and $\mathbf{q}=0.05(\mathbf{b}_1-\mathbf{b}_2)$. (d) Band structure of the heterostructure with $J_0=0.033t^0_\perp$, $\theta_0=\pi$ and $\mathbf{q}=0.05(\mathbf{b}_1-\mathbf{b}_2)$.
We took twist angle $1.44^\circ$ for (c), (d).
}
\label{fig1}
\end{figure}

Here we put forward helimagnetic encapsulation as a powerful tool to control flat bands in twisted graphene heterostructures.
In particular, we show that quasi-flat bands of twisted bilayer graphene (TBG) can be imprinted with strong spin textures
purely by exchange coupling with helimagnets.
Such spin textures 
do not rely on any kind of
spin-obit coupling effect in graphene.
We show the emergence of spin-textured bands for short and long helimagnetic
wavelengths and different commensuration with the
graphene sublattice structure, demonstrating the
robustness of helimagnet spin imprinting.
We finally discuss how helimagnet imprinting impacts possible 
superconducting states in the flat band systems.
Our results 
promote helimagnet encapsulation as a 
new strategy to design 
unconventional correlated states in twisted van der Waals heterostructures.

\section{Model}

We consider twisted bilayer graphene  at a twist angle $1.44^\circ$,
slightly above the flat band regime\cite{PhysRevLett.99.256802,PhysRevB.82.121407,Bistritzer2011}.
In this regime, the electronic structure of TBG shows
strongly renormalized Dirac cones\cite{PhysRevLett.99.256802} and isolated moire energy bands\cite{PhysRevB.82.121407,Bistritzer2011}. We consider TBG encapsulated between helimagnets
with in-plane magnetization as shown in Fig. \ref{fig1}(a). 
The impact of the helimagnet encapsulation is accounted by integrating out the degrees of freedom of the helimagnets, leading to an effective exchange field in the twisted graphene bilayer.
The effective Hamiltonian of the proximitized multilayer
takes the form

\beq \label{eq1}
\begin{aligned}
   H&=\sum_{l}\sum_{\langle i,j\rangle}\sum_{\alpha}tc^\dag_{i,\alpha,l}c_{j,\alpha,l}\\&+\sum_{l\neq l'}\sum_{i,j}\sum_{\alpha}t_\perp(i,j)c^\dag_{i,\alpha,l}c_{j,\alpha,l'}\\&+J\sum_l\sum_{i}\sum_{\alpha,\beta}\mathbf{M}_l(i)\cdot \boldsymbol{\sigma}_{\alpha\beta} c^\dag_{i,\alpha,l}c_{i,\beta,l}
\end{aligned}
\eeq
where $l=1,2$ is the layer index, $i,j$ are site indexes, and $\alpha,\beta$ are spin indexes. $t$ and $t_\perp$ are intra- and inter-layer hopping in TBG, $\langle i,j\rangle$ restricts the sum to nearest neighbours in the first term.
The interlayer hopping takes the form
 $
t_{\perp}(i,j) =
t^0_{\perp}
\frac{(z_i - z_j)^2 }{|\mathbf{r}_i - \mathbf{r}_j|^2}
e^{-\xi (|\mathbf{r}_i - \mathbf{r}_j|-d)}$\cite{PhysRevB.82.121407,PhysRevB.92.075402}, 
where $d$ is the distance
between layers and $\xi$ parameterizes the decay of the interlayer hopping\footnote{For computational efficiency we use a re-scaling relation\cite{PhysRevLett.119.107201,PhysRevB.98.195101,PhysRevB.101.060505} with $t^0_\perp=0.3t$}.
$J$ is the exchange coupling between the TBG and the helimagnets,  $\mathbf{M}_l(i)$ is the magnetization around site $i$ 
in the $l$th layer, and $\boldsymbol{\sigma}$ is a vector composed of Pauli matrices.
As a reference, the hopping constants for graphene are $t \approx 3$ eV and $t^0_\perp
\approx 500$ meV\cite{RevModPhys.81.109}.

The structure of the twisted multilayer is built as follows. We take $\mathbf{a}_{1,2}$ as lattice vectors of the bottom layer graphene: $\mathbf{a}_{1}=\sqrt{3}a(1,0)$ and $\mathbf{a}_{2}=\frac{\sqrt{3}a}{2}(1,-\sqrt{3})$, where $a$ is the lattice constant. The reciprocal vectors of graphene are then $\mathbf{n}_1=\frac{1}{3a}(\sqrt{3},1)$ and $\mathbf{n}_2=-\frac{2}{3a}(0,1)$. The moire superlattice vectors are then given by\cite{PhysRevB.92.075402}: $\mathbf{A}_{1}=(m+1)\mathbf{a}_{1}+m\mathbf{a}_{2}$ and $\mathbf{A}_{2}=(2m+1)\mathbf{a}_{1}-(m+1)\mathbf{a}_{2}$, where $m$ is an integer and in our case we take $m=11$. The reciprocal vectors of the moire superlattice are
$
\mathbf{b}_1=\frac{(m+1)\mathbf{n}_1+(2m+1)\mathbf{n}_2}{3m^2+3m+1}$ and $
\mathbf{b}_2=\frac{m\mathbf{n}_1-(m+1)\mathbf{n}_2}{3m^2+3m+1}
$.

We consider a general helimagnetic order that can be incommensurate with the graphene sublattice structure. 
For the sake of concreteness, 
we consider when the magnetization on sublattice $B$ has a relative rotation $\theta_0$ w.r.t. that on sublattice $A$
\beq\label{Eq:MAB}
\begin{aligned}
\mathbf{M}(i\in A)&=M_0(\cos(\mathbf{q}\cdot\mathbf{R}_i),\sin(\mathbf{q}\cdot\mathbf{R}_i),0)\\
\mathbf{M}(i\in B)&=M_0(\cos(\mathbf{q}\cdot\mathbf{R}_i+\theta_0),\sin(\mathbf{q}\cdot\mathbf{R}_i+\theta_0),0)
\end{aligned}
\eeq
with $M_0$ being the magnitude of the local magnetization, $\mathbf{q}$ being the characteristic wave vector of the helimagnet, and $\mathbf{R}_i$ being the coordinate of site $i$. Due to the superexchange mechanism\cite{PhysRev.79.350}, the local magnetization at top and bottom magnets are expected to align antiferromagnetically, so we consider $\mathbf{M}_{1}(i)=-\mathbf{M}_{2}(i)=\mathbf{M}(i)$. For the sake of simplicity, we consider $\mathbf{q}$ parallel to $\mathbf{b}_1-\mathbf{b}_2$, which is a high-symmetry direction of the moire Brillouin zone. When $\theta_0=0$, there is no sublattice imbalance and the magnets are fully characterized by the helimagnetic order (Fig. \ref{fig1}(b)).  When $\theta_0=\pi$, the helimagnetic order is 
overlayed with a staggered magnetization for the sublattices.

At low energy, the influence of the helimagnetization on the band structure of TBG depends only on the effective exchange field $J_0=JM_0$ and the ratio between the 
characteristic vector $\mathbf{q}$ and the moire periodicity. Given that the moire periodicity can be tuned by the twist angle $\theta$, we explore the following two regimes:
(i) the helimagnetization is commensurate with the TBG moire lattice, or (ii) the helimagnet periodicity is much smaller than the periodicity of the moire lattice. We show that in both cases the helimagnetization imprints a spin texture in quasi-flat bands in TBG. 

In the commensurate regime, 
the system can be solved with Eq. \eqref{eq1}. In the incommensurate regime, however, the whole system no longer has the periodicity of the moire lattice, and we adopt the generalized spinor-Bloch theorem to diagonalize the system\cite{Sandratskii1991,Egger2012,Braunecker2010}. We perform a local unitary transformation to the Hamiltonian such that the local magnetization is aligned along the $x$ direction for all sites\cite{Egger2012,Braunecker2010}:
\beq
U= \prod_i e^{-\frac{i}{2}\mathbf{q}\cdot\mathbf{r}_i\sigma_{z,i}}
\eeq

with $\sigma_{z,i}$ the spin Pauli matrix in site $i$. The transformed Hamiltonian becomes:

\beq \label{eq6}
\begin{aligned}
   H'&=U^\dag HU\\&=\sum_{l}\sum_{\langle i,j\rangle}\sum_{\alpha}te^{i\theta_{\alpha}(i,j)}c^\dag_{i,\alpha,l}c_{j,\alpha,l}\\&+\sum_{l\neq l'}\sum_{i,j}\sum_{\alpha}t_\perp(i,j)e^{i\theta_{\alpha}(i,j)}c^\dag_{i,\alpha,l}c_{j,\alpha,l'}\\&+J_0\sum_l\sum_{i\in \{A\}}\sum_{\alpha,\beta}f_l c^\dag_{i,\alpha,l}(\sigma_x)_{\alpha\beta}c_{i,\beta,l}\\&+J_0\sum_l\sum_{i\in \{B\}}\sum_{\alpha,\beta}\chi f_l c^\dag_{i,\alpha,l}(\sigma_x)_{\alpha\beta}c_{i,\beta,l}
\end{aligned}
\eeq
where $J_0=JM_0$ is the effective local exchange field,  $f_1=-f_2=1$, $\theta_{\uparrow}(i,j)=-\theta_{\downarrow}(i,j)=\mathbf{q}\cdot(\mathbf{R}_i-\mathbf{R}_j)$
and $\chi = \cos{\theta_0}$. Since the magnetization is uniform up to a sublattice imbalance in $H'$, $H'$ has the same periodicity as an isolated TBG. 
With no proximity effect, i.e. when $J_0=0$, the only difference between $H$ and $H'$ is the additional phases in the hopping terms, resulting in a momentum shift of $\pm\mathbf{q}/2$ for spin-up/down channels, respectively (Fig. \ref{fig1}(c)). We note that the additional phase in the first two terms correspond to
an artificial in-plane spin-orbit coupling, while the last two terms correspond to exchange terms. 
For finite $J_0$, spin-mixing occurs in the quasi-flat bands, creating a spin texture (Fig. \ref{fig1}(d)). In the following we address the spin-textured quasi-flat bands in TBG due to the proximity to the helimangets for both commensurate and incommensurate helimagnets.

 \begin{figure}[t!]
\center
\includegraphics[width=\linewidth]{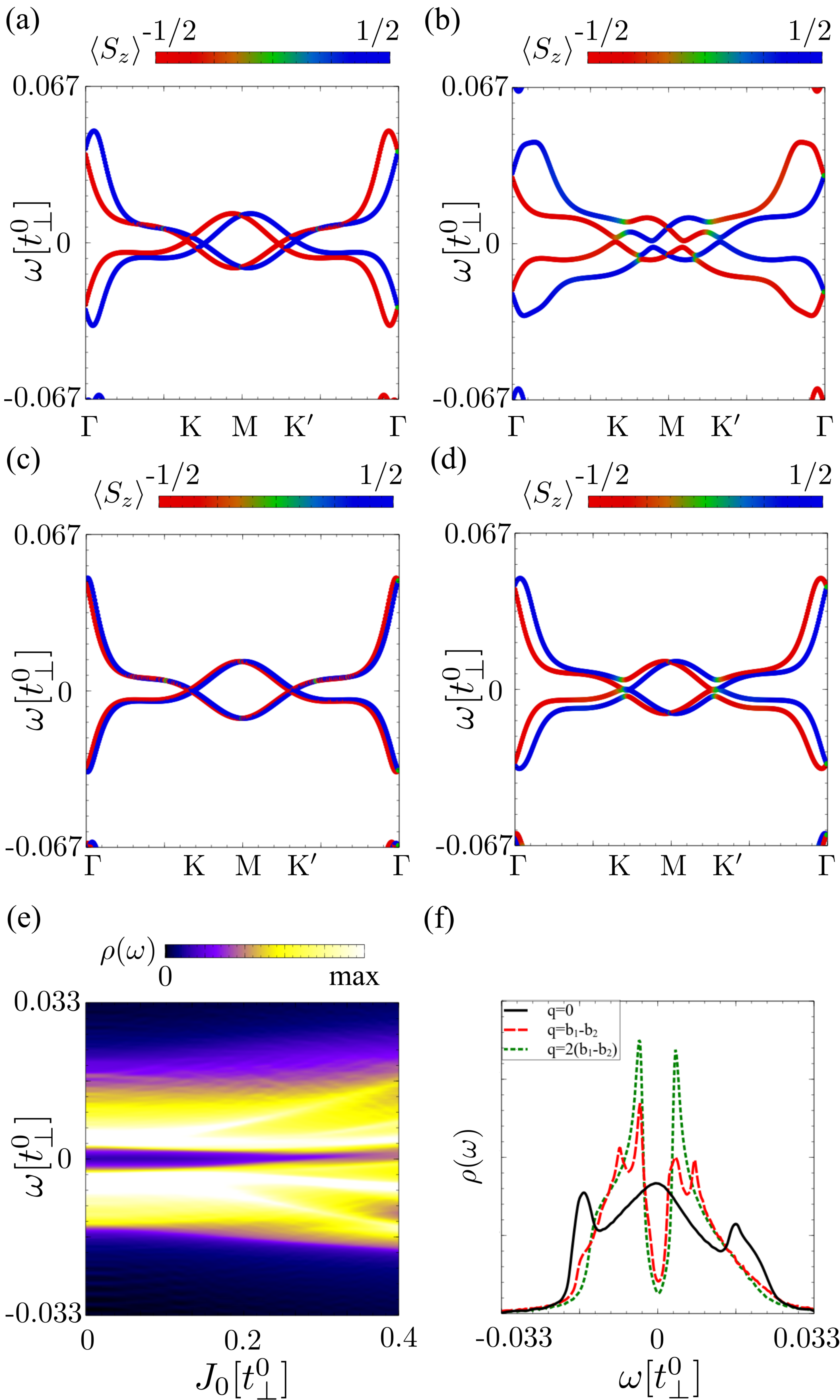}
\caption{ (a-d) Band structure and (e-f) DOS $\rho(\omega)$ of TBG with commensurate helimagnetic encapsulation Eq. \eqref{eq1}. The helimagnetic order, given by Eq. \eqref{Eq:MAB}, has a characteristic vector $\mathbf{q}$ commensurate with the TBG supercell and $\theta_0=0$. The effective local exchange is $J_0=JM_0$. We took in (a) $\mathbf{q}=\mathbf{b}_1-\mathbf{b}_2$ and $J_0=0.2t^0_\perp$, in (b) $\mathbf{q}=\mathbf{b}_1-\mathbf{b}_2$ and $J_0=0.4t^0_\perp$, in (c) $\mathbf{q}=2(\mathbf{b}_1-\mathbf{b}_2)$ and $J_0=0.2t^0_\perp$, in (d) $\mathbf{q}=2(\mathbf{b}_1-\mathbf{b}_2)$ and $J_0=0.4t^0_\perp$, in (e) $\mathbf{q}=\mathbf{b}_1-\mathbf{b}_2$ and in
(f) $J_0=0.2t^0_\perp$.}
\label{fig2}
\end{figure}

\section{Electronic structure with commensurate helimagnetic encapsulation} \label{sec3}
We start with the simplest case when the helimagnets are commensurate with the TBG supercell and the band structure can be computed with the Hamiltonian in Eq. \eqref{eq1}. We consider $J_0<t^0_\perp$, which corresponds to a realistic regime
with proximity to insulating magnets\cite{Leutenantsmeyer2016}. We note that in this regime, the Hamiltonian can be solved without including
a spin rotation in the Bloch phase, as the Hamiltonian in the original basis has the periodicity of the moire supercell.
We first comment on the case $\theta_0=\pi$. The spin spiral field acts as an artificial
spin-orbit coupling, generating a spin-splitting in momentum space. 
Interestingly, in this regime, the band structure does not exhibit 
sizable anticrossings driven by $J_0$. This behavior stems from the orthogonality between low energy states,
due to the $\pi$ Berry phase of the Dirac cones\cite{RevModPhys.81.109}. Since the low energy states are orthogonal, a coupling between these states does not result in a splitting. Therefore, the splitting, to the lowest order of $J_0$, comes from the coupling between the low energy bands with higher energy bands, and has a quadratic dependence on $J_0$. In the small $J_0$ regime we considered, a sizable gap will not open.

We now focus on the case of $\theta_0=0$, which leads to a strong change in the electronic structure induced by the helimagnetic field. The band structure along $\mathbf{b}_1-\mathbf{b}_2$ for different values of $\mathbf{q}$ and $J_0$ are shown in Fig. \ref{fig2}(a-d). We see that spin-splitting occurs in momentum space, stemming from the spiral field. In addition, the band structures exhibit anticrossings between the spin-channels for large $J_0$ (Fig. \ref{fig2}(b,d)), stemming from a first-order contribution in $J_0$ to the electronic energies. We note that both the spin-splitting (Fig. \ref{fig2}(a,c)) and the anticrossings (Fig. \ref{fig2}(b,d)) become smaller for larger $\mathbf{q}$. To investigate the impact of the helimagnets on the flatness of the quasi-flat bands, we present the density of states (DOS) $\rho(\omega)$ v.s $J_0$ for $\mathbf{q}=\mathbf{b}_1-\mathbf{b}_2$ (Fig. \ref{fig2}(e)). We see that as $J_0$ increases, the peaks in the DOS stemming from the quasi-flat bands split and the flat bands retain a small bandwidth. 
The previous phenomenology takes place for different values of $\mathbf{q}$ at $J_0=0.2t^0_\perp$ as shown in Fig. \ref{fig2}(f). We find that the DOS increases with $\mathbf{q}$, and for $\mathbf{q}=2(\mathbf{b}_1-\mathbf{b}_2)$ the DOS is almost the same as the DOS of pristine TBG. 
We have thus found that the helimagnets induce spin-splitting and anticrossing in quasi-flat bands in TBG. Both effects become more significant for larger effective exchange coupling $J_0$ and smaller characteristic vector $\mathbf{q}$ of the helimagnets. In the next section, we focus on the regime when $\mathbf{q}\ll\mathbf{b}_1-\mathbf{b}_2$, where similar effects are expected to occur at smaller values of $J_0$.

\section{Electronic structure with incommensurate helimagnetic encapsulation}
We now move on to investigate the modification of the band structure with helimagnetization in the small $\mathbf{q}$ regime. In particular, we focus on $\mathbf{q}\leq0.2(\mathbf{b}_1-\mathbf{b}_2)$ such that the helimagnetization does not result in intervalley scattering. We consider both $\theta_0=0$ and $\theta_0=\pi$ cases.

\begin{figure}[t!]
\center
\includegraphics[width=\linewidth]{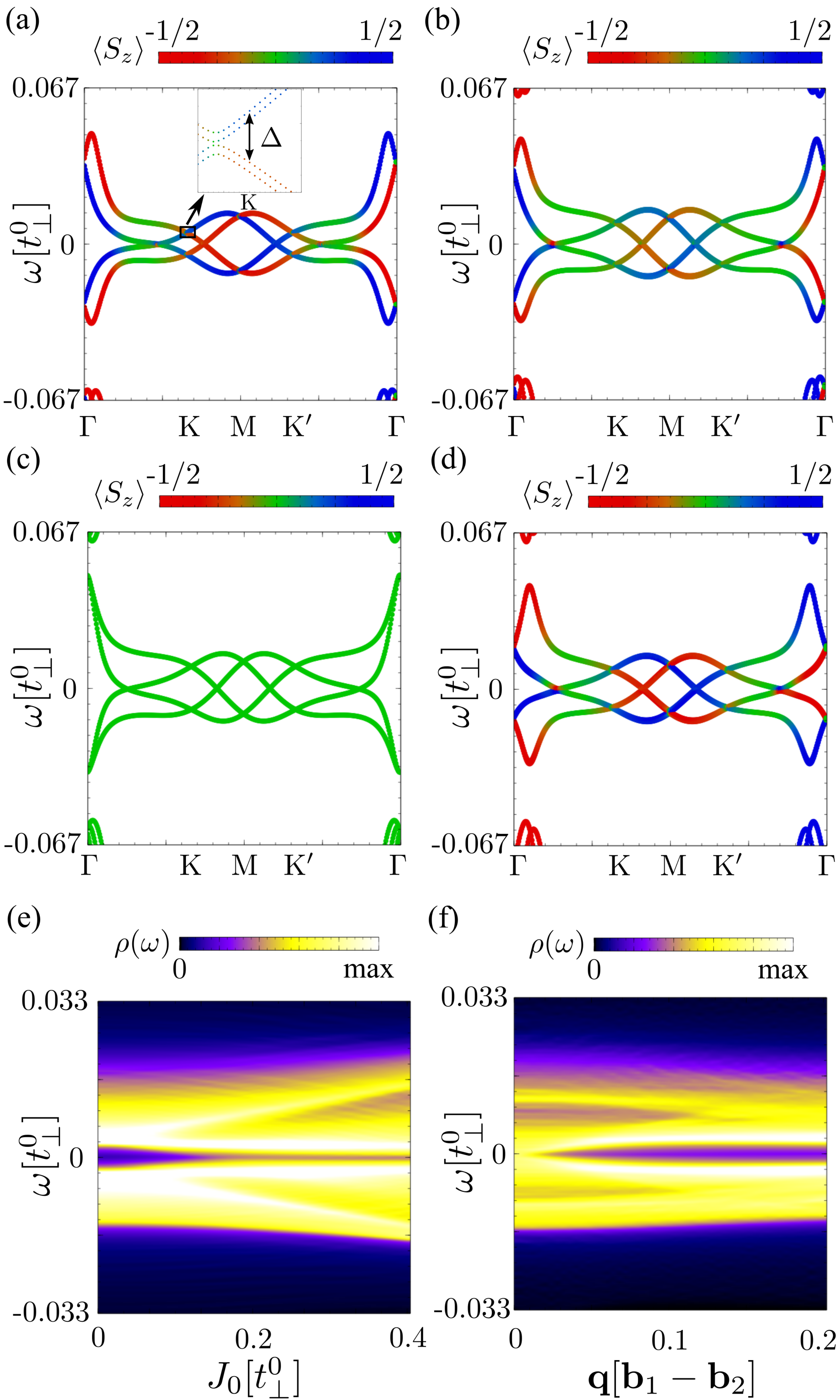}
\caption{(a-d) Band structure and (e-f) DOS $\rho(\omega)$ of TBG with incommensurate helimagnetic encapsulation Eq. \eqref{eq6}. The helimagnetic order, given by Eq. \eqref{Eq:MAB}, has a characteristic vector $\mathbf{q}$ much smaller than the moire reciprocal vectors of TBG and $\theta_0=0$. We took in (a) $\mathbf{q}=0.05(\mathbf{b}_1-\mathbf{b}_2)$ and $J_0=0.067t^0_\perp$. Inset: spin-splitting $\Delta$ at K point. We took in (b) $\mathbf{q}=0.05(\mathbf{b}_1-\mathbf{b}_2)$ and $J_0=0.13t^0_\perp$, in (c) $\mathbf{q}=0$ and $J_0=0.13t^0_\perp$, in (d) $\mathbf{q}=0.1(\mathbf{b}_1-\mathbf{b}_2)$ and $J_0=0.13t^0_\perp$, in (e) $\mathbf{q}=0.05(\mathbf{b}_1-\mathbf{b}_2)$ and in (f) $J_0=0.13t^0_\perp$.}
\label{fig3}
\end{figure}

 For $\theta_0=0$, we have a helimagnet that induces a nearly
 ferromagnetic exchange in neighboring sites. In this case, the band structure with $\mathbf{q}=0.05(\mathbf{b}_1-\mathbf{b}_2)$ and different values of $J_0$ is shown in Fig. \ref{fig3}(a,b). Similar to the large $\mathbf{q}$ case discussed in Sec. \ref{sec3}, spin-splittings and anticrossings appear in the
 nearly flat bands. The anticrossing gap $\Delta$ at the $K$ point as denoted in Fig. \ref{fig3}(a) exhibits a quadratic dependence on $J_0$, indicating that the anticrossing is caused by a second-order contribution. The reason that the first-order contribution does not cause spin-splitting is due to the orthogonality between the low energy eigenstates\cite{RevModPhys.81.109}, similar to the $\theta=\pi$ case in the commensurate $\mathbf{q}$ regime.
 Comparing the band structure with different $\mathbf{q}$ at $J_0=0.13t^0_\perp$ (Fig. \ref{fig3}(b-d)), we find that the spin texture exhibits strong dependence on $\mathbf{q}$, whereas the dispersion does not change substantially
 with $\mathbf{q}$. The DOS versus $J_0$ and $\mathbf{q}$ (Fig. \ref{fig3}(e,f)), show that helimagnetic encapsulation does not spoil the small bandwidth of the low energy bands. We find that as $J_0$ increases, the peaks in the DOS split and eventually the DOS slightly decreases. The DOS shows small $\mathbf{q}$-dependence for $\mathbf{q}>0.05(\mathbf{b}_1-\mathbf{b}_2)$, and only when $\mathbf{q}$ is close to 0 the DOS decreases. In this regime, a relatively small $J_0$ 
 leads to a strong spin texture in the quasi-flat bands, creating a strongly $\mathbf{q}$-dependent spin texture.

\begin{figure}[t!]
\center
\includegraphics[width=\linewidth]{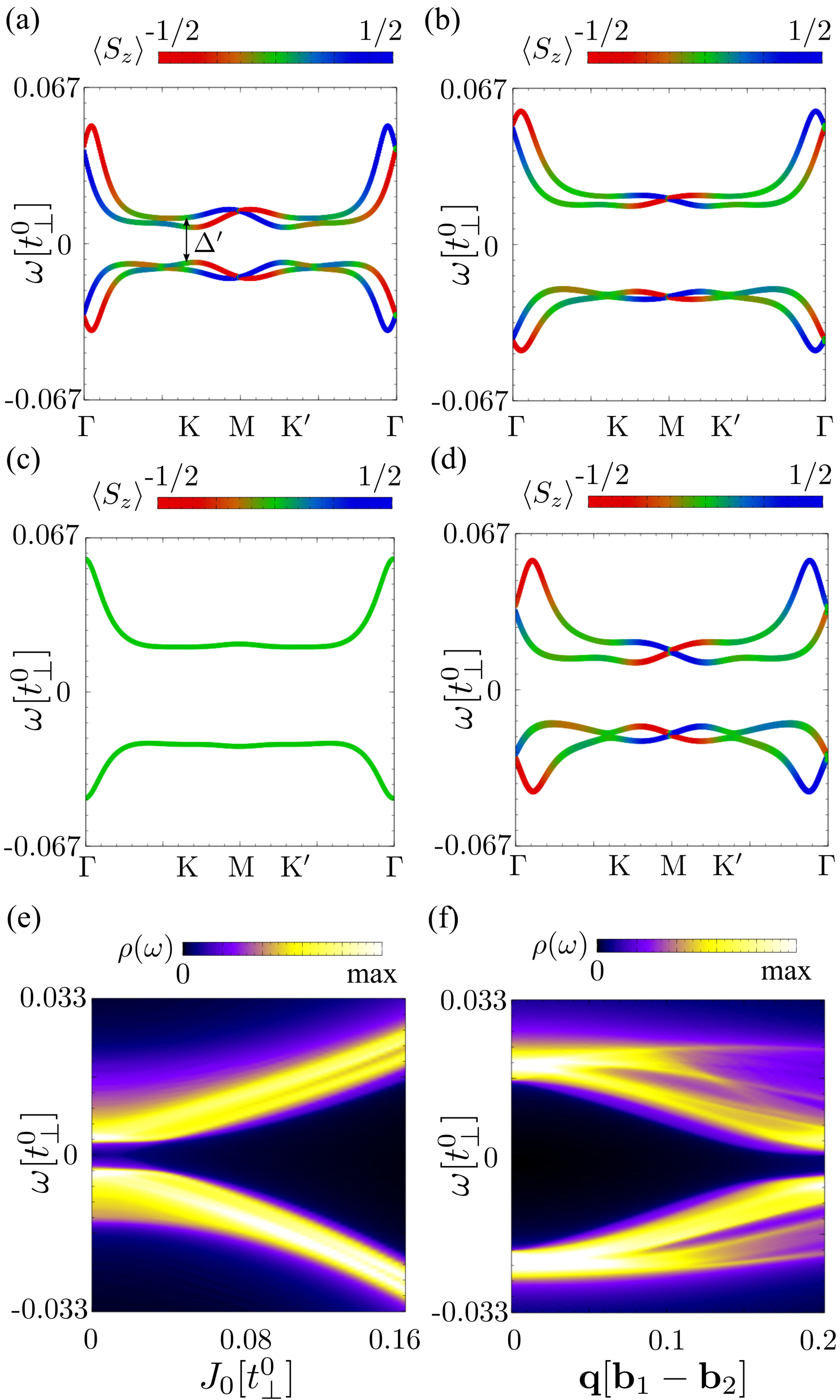}
\caption{(a-d) Band structure and (e-f) DOS $\rho(\omega)$ of TBG with incommensurate helimagnetic encapsulation Eq. \eqref{eq6}. The helimagnetic order, given by Eq. \eqref{Eq:MAB}, has a characteristic vector $\mathbf{q}$ much smaller than the moire reciprocal vectors of TBG and $\theta_0=\pi$. We took in (a) $\mathbf{q}=0.05(\mathbf{b}_1-\mathbf{b}_2)$ and $J_0=0.067t^0_\perp$, in (b) $\mathbf{q}=0.05(\mathbf{b}_1-\mathbf{b}_2)$ and $J_0=0.13t^0_\perp$, in (c) $\mathbf{q}=0$ and $J_0=0.13t^0_\perp$, in (d) $\mathbf{q}=0.1(\mathbf{b}_1-\mathbf{b}_2)$ and $J_0=0.13t^0_\perp$, in (e) $\mathbf{q}=0.05(\mathbf{b}_1-\mathbf{b}_2)$ and in (f) $J_0=0.13t^0_\perp$.}
\label{fig4}
\end{figure}

 We now move on to the case when $\theta_0=\pi$, i.e. the magnetization on the sublattices are opposite and we have a helimagnet inducing a nearly
 antiferromagnetic field in neighboring sites. The band structures with $\mathbf{q}=0.05(\mathbf{b}_1-\mathbf{b}_2)$ and different $J_0$ are shown in Fig. \ref{fig4}(a,b). In this regime, large anticrossings at $K$ and $K'$ appear,
 together with a simultaneous splitting of the Dirac cones. The splitting $\Delta'$ as denoted in Fig. \ref{fig4}(a) exhibits a linear dependence on $J_0$, indicating that the splitting stems from a first-order contribution.
 As a consequence of the splitting of the Dirac cones, the bands become flatter and shifted away from each other as $J_0$ increases, which is in contrast to the case when $\theta_0=0$. The $\mathbf{q}$-dependence of the bands with fixed $J_0=0.13t^0_\perp$ is shown in Fig. \ref{fig4}(b-d). Interestingly, both the spin texture and the dispersion depend strongly on $\mathbf{q}$. The low energy DOS versus $J_0$ and $\mathbf{q}$ is shown in Fig. \ref{fig4}(e,f),
 highlighting that the nearly flat bands maintain their flatness as the system develops a strong spin texture.
 
 Compared to the large $\mathbf{q}$ regime as discussed in Sec. \ref{sec3}, the small $\mathbf{q}$ regime with $\theta_0=0$ shows bands whose dispersion is less dependent on $\mathbf{q}$ whereas the spin texture is more sensitive on $\mathbf{q}$, making it more favorable for designing spin-textured flat bands. In addition, in the case when $\theta_0=\pi$, the quasi-flat bands in TBG can be further flattened with  helimagnetic encapsulation. In all the regimes discussed, quasi-flat bands with spin texture can be engineered in TBG by proximity to helimagnets. 

\begin{figure}[t!]
\center
\includegraphics[width=\linewidth]{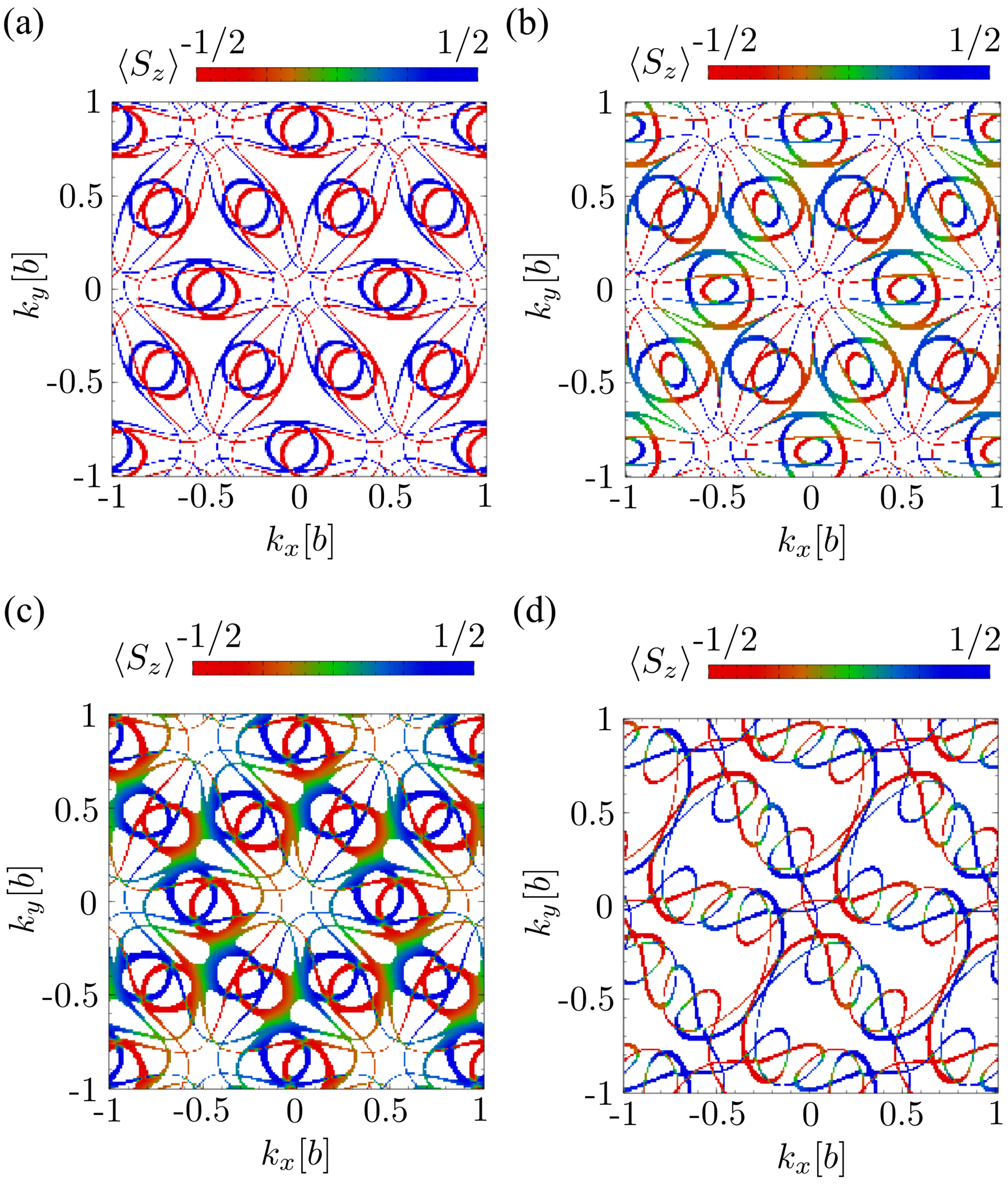}
\caption{Fermi surfaces of TBG with helimagnetic encapsulation with a doping of two holes per moire supercell, with different characteristic vector $\mathbf{q}$ and effective coupling $J_0$.
We took in (a) $\mathbf{q}=0.05(\mathbf{b}_1-\mathbf{b}_2)$ and $J_0=0$, in (b) $\mathbf{q}=0.05(\mathbf{b}_1-\mathbf{b}_2)$, $J_0=0.067t^0_\perp$ and $\theta_0=0$, in (c) $\mathbf{q}=0.05(\mathbf{b}_1-\mathbf{b}_2)$, $J_0=0.067t^0_\perp$ and $\theta_0=\pi$, and in (d) $\mathbf{q}=\mathbf{b}_1-\mathbf{b}_2$, $J_0=0.2t^0_\perp$ and $\theta_0=0$. $b=|\mathbf{b}_1|=|\mathbf{b}_2|$ is the length of the reciprocal vectors.}
\label{fig5}
\end{figure}

\section{Spin-textured Fermi surfaces and superconducting instabilities}

In this section we discuss the impact of the imprinted spin texture on the superconducting instability in TBG. Superconductivity in the weak coupling limit is understood in terms of a Fermi surface instability by the formation of Cooper pairs with total momentum equal to zero\cite{RevModPhys.63.239}. This means that the electrons forming the pairs have opposite momenta, and therefore belong to states that are related by inversion or by time-reversal symmetry (TRS). The breaking of these two key symmetries by the helimagnet encapsulation in principle hinders the formation of Cooper pairs. 
In the following we show explicitly the spin-textured Fermi surfaces of the helimagnetically-encapsulated TBG and discuss their impact on the robustness of different types of superconducting states. We consider a hole doping with two holes per moire supercell.

In the absence of exchange coupling to the helimagnets, the Fermi surfaces have six-fold symmetry around the center of the Brillouin zone and are spin-degenerate, such that all types of spin-configuration for the Cooper pairs are possible. As a reference, in the absence of the effective exchange magnetic field, the Fermi surface of the Hamiltonian
in the rotated basis Eq. \eqref{eq6} with $\mathbf{q}=0.05(\mathbf{b}_1-\mathbf{b}_2)$ is shown in Fig. \ref{fig5}(a), where we observe the relative shifting of spin-up/down channels in reciprocal space, reminiscent of a Rashba spin-orbit coupling.
Once $J_0\neq0$, the Fermi surfaces for both $\theta_0=0$ and $\theta_0=\pi$ exhibit spin-mixing
(Fig. \ref{fig5}(b,c)). 
Even though TRS is now explicitly broken, the Fermi surface still displays states related by time-reversal symmetry. This can be understood by the combination of TRS with a horizontal mirror symmetry \cite{Fischer:2018}. Under these circumstances,  spin-singlet ($\sim |\uparrow\downarrow\rangle - |\downarrow\uparrow\rangle$)  and spin-triplet pairing with the d-vector along the z-axis ($\sim |\uparrow\downarrow\rangle + |\downarrow\uparrow\rangle$) are stable\cite{Anderson1959,PhysRevB.30.4000,Andersen2020,PhysRevB.98.024501,PhysRevB.94.104501}. The spin-mixing is also present in the Fermi surface when a commensurate helimagnet encapsulation is considered (Fig. \ref{fig5}(d)), and the above argument applies. Interestingly, as the proximity to the helimagnet introduces inversion symmetry breaking, the mixing of superconducting states of different parity is allowed\cite{Smidman2017,PhysRevResearch.3.033133}. If the dominant pairing channel is $s$-wave, inversion symmetry breaking would induce a $p$-wave component to the superconducting gap structure due to the presence of an effective Rashba spin-orbit coupling. The converse situation is also true, if the dominant pairing channel is $p$-wave, some $s$-wave component is induced by inversion symmetry breaking. Ultimately, this allows to change the parity of the dominant channel in the superconducting gap by encapsulation with strong helimagnets.

\section{Conclusion}

To summarize, we have shown how nearly-flat bands of twisted graphene multilayers
can be imprinted with non-trivial spin textures via helimagnet encapsulation.
In particular, we have shown the emergence of strong
spin textures both in the limit of large and small q-vector in comparison with the
moire unit cell. Interestingly, we demonstrated that the magnetic encapsulation
alone is capable of opening up a gap at the Dirac points of the flat bands,
maintaining the flatness of the low energy bands.
We have shown that the existence of such spin textures dramatically impacts
the spin structure of the Fermi surface. Finally, we discussed
the potential of helimagnetic encapsulation to
promote unconventional superconducting states
in the system.
Our results put forward helimagnet encapsulation as a powerful
technique to design spin-textured flat bands and promote
unconventional superconducting states
in twisted graphene multilayers.

\begin{acknowledgments}
We acknowledge
the computational resources provided by
the Aalto Science-IT project,
and the
financial support from the
Academy of Finland Projects No.
331342 and No. 336243. AR acknowledges the support
from an Ambizione grant by the Swiss National Science Foundation.
We thank P. Liljeroth, S. Kezilebieke, A. Fumega
and P. Törmä for useful discussions.
\end{acknowledgments}

\bibliography{main}

\end{document}